\documentclass[pdflatex,sn-basic,iicol,sort&compress]{sn-jnl}%

\usepackage{graphicx}
\usepackage{wrapfig,lipsum}
\usepackage{amssymb}
\usepackage{amsmath}
\usepackage{afterpage}
\usepackage{xcolor}
\usepackage{booktabs}
\usepackage{fontawesome}
\setcitestyle{numbers,square,citesep={,}} 

\usepackage{multicol,cuted}
\usepackage{ulem}
\usepackage{geometry}

\definecolor{rosso}{cmyk}{0,1,1,0.4}
\definecolor{rossos}{cmyk}{0,1,1,0.6}
\definecolor{rossoc}{cmyk}{0,1,1,0.2}
\definecolor{blu}{cmyk}{1,1,0,0.3}
\definecolor{blus}{cmyk}{1,1,0,0.3}
\definecolor{bluc}{cmyk}{1,1,0,0.1}
\definecolor{verde}{cmyk}{0.92,0,0.59,0.25}
\definecolor{verdec}{cmyk}{0.92,0,0.59,0.15}
\definecolor{verdes}{cmyk}{0.92,0,0.59,0.4}
\hypersetup{colorlinks,bookmarksopen,bookmarksnumbered,
linkcolor=blus,pdfstartview=FitH,urlcolor=rossos,citecolor=verdes}

\newcommand{\be}{\begin{equation}}
\newcommand{\ee}{\end{equation}} 
\newcommand{\bry}{\begin{array}}
\newcommand{\ery}{\end{array}} 
\newcommand{\dst}{\displaystyle} 
\newcommand{\bit}{\begin{itemize}} 
\newcommand{\eit}{\end{itemize}} 
\newcommand{\ben}{\begin{enumerate}} 
\newcommand{\een}{\end{enumerate}}

\begin{document}


\title[\centering More variables or more bins? \\ Impact on the EFT interpretation \\ of Drell-Yan measurements]{\centering More variables or more bins? \\ Impact on the EFT interpretation \\ of Drell-Yan measurements}

\author*[1,2]{\fnm{Samuele} \sur{Grossi}}\email{samuele.grossi@ge.infn.it}
\author*[2]{\fnm{Riccardo} \sur{Torre}}\email{riccardo.torre@ge.infn.it}

\affil*[1]{\centering\orgdiv{Department of Physics}, \orgname{University of Genova}, \orgaddress{\street{Via Dodecaneso 33}, \\\city{Genova}, \postcode{16146}, \country{Italy}}}
\affil*[2]{\orgdiv{INFN}, \orgname{Sezione di Genova}, \orgaddress{\street{Via Dodecaneso 33}, \city{Genova}, \postcode{16146}, \country{Italy}}}

\abstract{We generalize previous studies on constraining operators of the Standard Model Effective Field Theory using Drell-Yan (DY) measurements to include at the same time all relevant operators and uncertainties. It has been shown that fully differential measurements (triple differential for neutral and double differential for charged) are more sensitive to EFT effects. Nevertheless, due to the finite statistics, the fully differential measurements sacrifice some statistical power on the shape (less invariant mass or transverse momentum bins) in favour of more kinematic variables. We show that when the observables are particularly sensitive to the shape of the distributions, such as the invariant mass of the two leptons in neutral DY, the single differential measurement with more bins, may be as sensitive as the fully differential one, at least for specific EFT operators. This suggests to always supplement fully differential analyses with projections into the relevant distributions evaluated with finer bins.}

\newgeometry{textwidth=5.5in, left=1in, top=0.5in}

\maketitle
\thispagestyle{empty}
\clearpage

\restoregeometry
\tableofcontents

\section{Introduction}
Our current best chance to investigate New Physics (NP) scenarios responsible for explaining the Electroweak (EW) hierarchy hinges on particle colliders, which, with their 
high energy and luminosity, are able to probe, both directly and indirectly, the existence of new particles and interactions. In all scenarios where new particles are too heavy to be directly produced at the LHC, the only way to detect their presence is through the indirect effects they have on Standard Model (SM) observables. The Standard Model Effective Field Theory (SMEFT)~\cite{Brivio:2017vri,Cohen:2020xca,Falkowski:2019tft,deBlas:2017xtg,McCullough:2023szd} is a fundamental tool in this context, since it helps to organize the possible NP effects of SM observables systematically. In this regard, the SMEFT can be seen as the modern version of the SM, of which the dimension-$4$ SM Lagrangian is the long distance description. 

Hadron colliders have long been recognized as challenging for precision physics due to the inherently noisy conditions of hadronic interactions, which significantly complicate precise measurements. Nevertheless, in the last decade, the LHC ability to pursue a high-energy precision program, has emerged. This program complements the Higgs precision program on one side, and the ``low-energy'' flavor program on the other side, showcasing the versatility of hadron colliders in advancing diverse research objectives, beyond the traditional search for new particles at the energy frontier. 

The so-called LHC high-$p_T$ precision program of the LHC mainly relies on the presence of SMEFT operators whose effects on precision observables grow with energy. A prototype example, within this program, is the Drell-Yan (DY) process, which is experimentally clean and theoretically well-understood, and is sensitive, in a generic SMEFT basis, to a set of two-quark-two-lepton four-fermion operators, and to a set of operators that modify the properties of the SM gauge bosons that contribute to the DY process in the SM. 

The DY process has originally been considered in this framework in Ref.~\cite{Farina:2016rws}, where it was shown how, exploiting the growing-with-energy effects of some relevant EFT operators on measurable observables, could lead to EW precision measurements at the LHC that could compete, or even surpass, the ones obtained at LEP. Studies targeting DY have been improved \cite{Farina:2016rws,Torre:2020aiz,Panico:2021vav}, even with a first attempt of interpretation of a NP search in terms of EFT operators by the CMS Collaboration \cite{CMS:2022krd}. The same idea of targeting growing with energy effects to indirectly explore heavy NP was also applied to a variety of other channels, among which di-jets \cite{Alioli:2017nzr,Alioli:2017jdo,Han:2023krp,Ethier:2021bye}, di-bosons \cite{Giudice:2007fh,Liu:2016idz,LHCHiggsCrossSectionWorkingGroup:2016ypw,CMS:2018uag,Rossia:2023hen,Degrande:2023iob,ATLAS:2024lyh,Englert:2023qun}, di-quarks \cite{Alioli:2017jdo}, and di-tops \cite{Ellis:2020unq,CMS:2020lrr,Tong:2023lms,Elmer:2023wtr}, and even four-tops \cite{Englert:2019zmt}.

In this work we focus again our attention on the DY processes. Our aim is to generalize previous constraints on the Wilson coefficient of SMEFT operators, including at the same time all the relevant operators~\cite{Panico:2021vav} and all the relevant uncertainties~\cite{Torre:2020aiz}. Moreover, we investigate the impact of different binning specifications and single vs multi-differential measurements to assess the best strategy of presenting measurements to maximize the sensitivity to SMEFT operators.

We consider the dimension-$6$ two-quark-two-lepton four-fermion operators as written in the Warsaw basis~\cite{Grzadkowski:2010es}, listed in Table~\ref{tab:EFToperators}. This represents, in the chosen basis, the full set of operators that lead to growing-with-energy effects in the DY process \cite{Panico:2021vav}. In our analysis we consider both the neutral DY, with a di-lepton final state, and the charged DY, with a lepton and a neutrino final state, resulting experimentally in a single lepton plus missing energy final state. We exploit both the fully differential cross-section, triple differential for the neutral channel, and double differential for the charged one, and the single differential cross-section, in the di-lepton invariant mass $m_{\ell\ell}$ (neutral) and lepton transverse momentum $p_T$ (charged), respectively. 

\begin{table}[t!]
    \centering
    \begin{tabular}{|l|}
    \hline
     Dimension-6 current-current operators\\
     \hline
     $O^{(3)}_{lq}=(\overline{l}_L\sigma_I\gamma^\mu l_L)(\overline{q}_L\sigma_I\gamma_{\mu}q_L)$\\
     $O^{(1)}_{lq}=(\overline{l}_L\gamma^\mu l_L)(\overline{q}_L\gamma_{\mu}q_L)$\\
     $O_{qe}=(\overline{q}_L\gamma^{\mu}q_L)(\overline{\ell}_R\gamma_{\mu}\ell_R)$\\
     $O_{lu}=(\overline{l}_L\gamma^\mu l_L)(\overline{\texttt{u}}_R\gamma_{\mu}\texttt{u}_R)$\\
     $O_{ld}=(\overline{l}_L\gamma^\mu l_L)(\overline{\texttt{d}}_R\gamma_{\mu}\texttt{d}_R)$\\
     $O_{eu}=(\overline{\ell}_R\gamma^\mu\ell_R)(\overline{\texttt{u}}_R\gamma_{\mu}\texttt{u}_R)$\\
     $O_{ed}=(\overline{\ell}_R\gamma^\mu\ell_R)(\overline{\texttt{d}}_R\gamma_{\mu}\texttt{d}_R)$\\ \hline
    \end{tabular}\vspace{2mm}
    \caption{The seven dimension-6 contact operators contributing to Drell-Yan processes written in the notation of ref.~\cite{Grzadkowski:2010es}.}
    \label{tab:EFToperators}
\end{table}

Since no new experimental measurements of differential DY at high energy has appeared since the last analysis of Ref.~\cite{Panico:2021vav}, we rely, in our analysis, on simulated data. This allows us to explore the impact of different binning strategies, and to assess the importance of presenting independent single and multi-differential measurements. We do so by comparing the sensitivity of analyses based on the fully-differential cross-section, on the single differential cross-section obtained by integrating the fully-differential one over the angular variables, and the ``enhanced" single differential cross-section obtained with a finer binning in the dimensionful kinematic variables, that are the invariant mass of the two leptons in the neutral DY, and the lepton transverse momentum in the charged DY.

We find that, for parameters that are particularly sensitive to the shape of the distribution in the dimensionful kinematic variables, the single differential cross-section in which the full statistics is used to optimize the binning compatibly with the statistical uncertainty, can be as sensitive as the fully differential one. This suggests that, in forthcoming experimental analyses, it would be optimal to have available both fully-differential and single differential (with optimized binning) cross-section information.

\section{Cross-sections parametrization}
The SM+EFT cross-section is obtained using the reweighting strategy introduced in Ref.~\cite{Torre:2020aiz}. This is based on the fact that in DY the new physics contributions factorize not only with respect to the tree-level cross-section, but also with respect to QCD radiative corrections. Such factorization holds separately for each chirality channel, and allows one to generate events for the full SM+EFT process, for any value of the Wilson coefficients, and at NLO QCD accuracy, by just reweighting events for the SM process at NLO QCD accuracy. The reweighting coefficients depend on the chirality of the quarks and leptons and on kinematic quantities such as the partonic center-of-mass squared $s$. The parton shower at NLO QCD accuracy is not modified by the EFT operators, therefore there is no need to account for their contribution. The SM cross-section and the parton shower at NLO QCD accuracy have been obtained using respectively \textsc{Powheg}~\cite{Alioli:2008gx,Frixione:2007vw} and \textsc{pythia} 8~\cite{Sjostrand:2014zea}. To obtain results whose precision matches the experimental measurements, also the NNLO QCD corrections have to be taken into account for the Standard Model contributions (using FEWZ~\cite{Li:2012wna}). These contributions are completely negligible for the new physics. Finally, the reweighting coefficients can be modified in order to include also the EW next-to-Leading-Log (NLL) corrections which become important in the high energy regime we are interested in \cite{Torre:2020aiz}.

All relevant uncertainties have been taken into account following the prescription of Ref.~\cite{Torre:2020aiz}. Exploiting the fact that the cross-section in each bin, in both the neutral and the charged channels, is a quadratic polynomial in the Wilson coefficients, we can perform the Cholesky decomposition with coefficients parametrizing the SM contribution, the interference between SM and EFT operators, and the quadratic EFT contributions. In this way we obtain a total of $36$ weights for the neutral channel ($1$ for the SM, $7$ for the interference, and $28$ for quadratic terms) and $3$ weights ($1$ SM, $1$ interference, and $1$ quadratic) for the charged one (see Appendix~\ref{app:cross_section} for details). All uncertainties are parametrized as fluctuations of these weights, which we refer to as Cholesky coefficients. 

\section{Uncertainties} 
In order to match the high precision measurements with theoretical predictions one has to take into account the most important theoretical and experimental sources of uncertainties. This is done through the introduction of nuisance parameters as in Ref.~\cite{Torre:2020aiz}. We assume that the nuisance parameters deriving from theoretical uncertainties modify the Cholesky coefficients (in each bin), while the ones linked to the experimental uncertainties have a direct impact on the number of expected events in each bin. We discuss the implementation of both classes of uncertainties in the following. To do so we indicate with $\overline{c}_{i, I}$ the $i$-th Cholesky coefficient in the $I$-th bin, calculated with Standard Model Central values of $\alpha_{\text{S}}$, Parton Density Functions (PDFs), and factorization and renormalization scales that we specify below, and with $c_{i,I}$ the corresponding Cholesky coefficient as function of the nuisance parameters. For simplicity, we do not separate the discussion between neutral and charged channels. It should nevertheless be clear that all quantities differ in each channel.
\begin{itemize}
    \item {\bf Theory uncertainties}
    \begin{itemize}
        \item {\it Monte Carlo statistic}\\
        The uncertainty deriving from Monte Carlo statistics is negligible if the simulations provide accurate enough predictions for the SM terms, well below 1$\%$. This is guaranteed by the fact that the new physics contributions are accounted for using reweighting, so that their accuracy aligns with that of the SM terms. \vspace{1mm}
        \item {\it Strong coupling constant}\\
        The uncertainty associated to the value of $\alpha_{\text{S}}$ is accounted for through a single nuisance parameter $\theta^{\alpha_{\text{S}}}$, which is the same across all channels and bins. The effect of $\theta^{\alpha_{\text{S}}}$ is estimated using \textsc{Powheg} SM DY~\cite{Alioli:2008gx} Monte Carlo samples reweighted for upper ($\alpha^{\text{u}}_{\text{S}}=0.1195$), lower ($\alpha^{\text{l}}_{\text{S}}=0.1165$), and central value ($\alpha_{\text{S}}=0.1180$) of $\alpha_{\text{S}}$ at the scale of the $Z$ mass.
        Since this uncertainty is not the leading one in the SM, we can ignore its effect on the new physics Cholesky coefficients and only retain the SM part, parameterized by the coefficient $c_0$:
        \begin{equation}
            \small
            \begin{array}{l}
            \dst c_{0,I}(\theta^{\alpha_{\text{S}}})=\overline{c}_{0,I}e^{k^{\alpha_{\text{S}}}_I\theta^{\alpha_{\text{S}}}}=e^{k^{\alpha_{\text{S}}}_I\theta^{\alpha_{\text{S}}}}\,, \vspace{2mm}\\
            \dst k^{\alpha_{\text{S}}}_I\hspace{-0.5mm}=\hspace{-0.5mm}\text{max}\left(\mid \hspace{-0.5mm} c_{0,I}^\text{u}-\overline{c}_{0,I}\hspace{-0.5mm}\mid,\mid\hspace{-0.5mm} c_{0,I}^\text{l}-\overline{c}_{0,I}\hspace{-1mm}\mid \right)\,.
            \end{array}
            \normalsize
        \end{equation}
        with $c_{0,I}^\text{l}=c_{0,I}(\alpha_{\text{S}}^\text{l})$ and $c_{0,I}^\text{u}=c_{0,I}(\alpha_{\text{S}}^\text{u})$. \vspace{1mm}
        \item {\it Missing higher orders (QCD and EW)}\\
        The uncertainty deriving from the truncation of the perturbative QCD series are accounted for by the introducion of a nuisance parameter $\theta_I^{\text{TU}}$ for each bin. We consider different values of the factorization and renormalization scales, $\mu_F$ and $\mu_R$, respectively: their central values are set to $\overline{\mu}_R=\overline{\mu}_F=\sqrt{\hat{s}}$ and we let them vary independently by multiplicative factors $2^{\pm1}$, $2^{\pm\frac{1}{2}}$, and $1$, with the latter value corresponding to the central value. This gives a grid with 25 values.\\
        Again, the missing higher order uncertainty is not leading in the SM. In particular, while the contribution deriving from the truncation of the NLO EW perturbative series could be completely neglected, the one linked to the truncation of the QCD NNLO perturbative series is relevant only for the $c_0$ SM coefficient. Such contribution is parameterized as:
        \begin{equation}
            \small
            \begin{array}{l}
            \dst c_{0,I}(\theta_I^{\text{TU}})=\overline{c}_{0,I}e^{{k_I^{\theta^{\text{TU}}}\theta_I^{\text{TU}}}}=e^{{k_I^{\theta^{\text{TU}}}\theta_I^{\text{TU}}}}\,, \vspace{2mm}\\
            \dst k_I^{\theta^{\text{TU}}}\hspace{-0.5mm}=\hspace{-0.5mm}\text{max}\left(\frac{\mid c^{\text{max}}_{0,I}-\overline{c}_{0,I}\mid}{10},\frac{\mid c^{\text{min}}_{0,I}-\overline{c}_{0,I}\mid}{10}\right)\,,
            \end{array}
            \normalsize
        \end{equation}
        where $c_{0,I}^{\text{max}}$ and $c_{0,I}^{\text{min}}$ are the maximum and minimum value of $c_{0,I}$ within the 25 different replicas specified above. \vspace{1mm}
        \item {\it Parton Distribution Functions}\\
        PDF uncertainty is the most important theoretical uncertainty in the SM DY process \cite{Gao:2017yyd,Hammou:2023heg,Amoroso:2022eow}. Therefore, we account for it in the Cholesky coefficients of both the SM and the new physics contributions. 
        The PDF uncertainties are parametrized by a vector of nuisance parameter $\theta^{\text{PDF}}_i$, corresponding to the eigenvalues of the PDFs within the Hessian representation, for each bin. As before, we use \textsc{Powheg} to get the weights of the different Hessian components in the SM calculation. In our case it is enough to consider the $30$ components in the  PDF set PDF4LHC15$\_$\textsc{nlo}$\_$30$\_$\textsc{pdfas} (code 90400 in the LHAPDF database~\cite{Buckley:2014ana}). The advantage of the Hessian set is that it automatically provides a definition of the relevant nuisance parameters that can be used across different processes, simplifying the combination of different channels. The parametrization of the Cholesky coefficients as function of the PDFs nuisance parameters is:
        \begin{equation}\label{eq:ckPDF}
            \small
            \begin{array}{l}
            \dst c_{k,I}(\theta^{\text{PDF}}_i)=\overline{c}_{k,I}\exp\left[\sum_{i=1}^{30}\frac{c_{k,I}^{(i)}-\overline{c}_{k,I}}{\overline{c}_{k,I}}\theta_i^{\text{PDF}}\right]\,, \vspace{2mm}\\
            \dst \text{for}\,\, k=0,1,...,35\qquad (\text{Neutral channel}), \vspace{2mm}\\
            \dst \text{for}\,\, k=0,1,2\qquad\quad\,\, (\text{Charged channels}). 
            \normalsize
            \end{array}
        \end{equation}
    \end{itemize}    
    \item {\bf Experimental uncertainties}\\
        Uncertainties associated to the experimental setup could only be accounted for within an analysis or with available information from the experiments. Despite our code gives us full flexibility in accounting for experimental systematic uncertainties with full correlation information, possibly different in the electron and muon channels, in this work we considered a relatively simple pattern of uncertainties, consistent with past analyses. We assume a $2\%$ uncorrelated uncertainty, parametrized by a single nuisance parameters $\theta^{\text{L}}$, across all bins and channels, from the determination of the integrated luminosity. For all the other experimental systematic uncertainties, we consider $2\%$ and $5\%$ uncertainties in each bin, uncorrelated among bins, for the neutral and charged channels, respectively. These are parametrized by a set of nuisance parameters $\theta_{I}^{\text{exp}}$ for each bin and each channel.

        We assume that the experimental uncertainties do not affect directly the Cholesky coefficients, and only modify the number of expected events in each bin $\mu_{I}$ from its theoretical prediction $\mu^{\text{th}}_I$ as follows: 
        \begin{equation}
        \mu_I=\mu^{\text{th}}_I\exp\left(\sum_J\left[\sqrt{\Sigma^{\text{exp}}}\right]^J_I\theta^{\text{exp}}_J+0.02\theta^{\text{L}}\right)\,. \label{eq:expevents}
        \end{equation}
        Here $\Sigma^{\text{exp}}$ is the covariance matrix of the experimental systematic uncertainties in the space of bins, that we take proportional to the identity matrix, and $\mu^{\text{th}}_I=L\cdot\sigma^{\text{th}}_I$, where $L$ is the integrated luminosity, and $\sigma^{\text{th}}_I$ is the cross-section in each bin including all sources of theoretical uncertainties specified before.
\end{itemize}

\section{The likelihood}
The constraints on the Wilson coefficients are obtained using the profiled likelihood ratio test, with the test-statistic $t_{\boldsymbol{\mu}}$ defined by
\begin{equation}
    t_{\boldsymbol{\mu}}\hspace{-0.5mm}=\hspace{-0.5mm}-2\hspace{-0.5mm}\left(\text{sup}_{\boldsymbol{\delta}}\log \mathscr{L}(\boldsymbol{\mu}, \boldsymbol{\delta})\hspace{-0.5mm}-\hspace{-0.5mm}\text{sup}_{(\boldsymbol{\mu},\boldsymbol{\delta})}\log \mathscr{L}(\boldsymbol{\mu}, \boldsymbol{\delta})\hspace{-0.5mm}\right)\,. \label{eq:tmu}
\end{equation}
Here $\mathscr{L}$ is the likelihood as function of the parameters, $\boldsymbol{\mu}$ represents the parameters of interest, that are the seven Wilson coefficients corresponding to the EFT operators in Table~\ref{tab:EFToperators}, and $\boldsymbol{\delta}$ are the nuisance parameters.\\
Assuming that $t_{\boldsymbol{\mu}}$ follows asymptotically a $\chi^2$ distribution with a number of degrees of freedom equal to the number of parameters of interest \cite{Cowan:2010js}, we can set confidence level boundaries on the Wilson coefficients. Bin by bin the likelihood is a Poisson distribution of the number of observed events $n_I$ with mean $\mu_I$, multiplied by the likelihood that constrains each nuisance parameter (with auxiliary data), which can be parametrized as a standard Normal distribution, since the relevant scales have already been taken into account in defining the dependence of the Cholesky coefficients on the nuisance parameters. The combined likelihood is then simply written as:
\begin{equation}\label{eq:comblikelihood}
\begin{array}{lll}
    \mathscr{L}_{\text{comb}}(\boldsymbol{\mu},\boldsymbol{\delta}) &=&\mathscr{L}_{\text{n}}(\boldsymbol{\mu},\boldsymbol{\delta}_{\text{n}})\times\mathscr{L}_{\text{c},+}(\boldsymbol{\mu},\boldsymbol{\delta}_{\text{c}})\vspace{2mm}\\
    &\times&\mathscr{L}_{\text{c},-}(\boldsymbol{\mu},\boldsymbol{\delta}_{\text{c}})\times\mathscr{L}_{\text{aux}}(\boldsymbol{\delta})\,,
\end{array}
\end{equation}
where the explicit definition of each term is given in Appendix~\ref{likelihood}. 
\begin{table*}[t!]
    \centering
    \resizebox{\linewidth}{!}{
    \begin{tabular}{|c|c|c|c|c|c|c|c|c|c|}
    \hline \textbf{95\% CL} & \multicolumn{3}{c|}{$\mathcal{L}=100\,\text{fb}^{-1}$} & \multicolumn{3}{c|}{$\mathcal{L}=300\,\text{fb}^{-1}$} & \multicolumn{3}{c|}{$\mathcal{L}=3000\,\text{fb}^{-1}$} \\ \cline{2-10}
    $[10^{-9} \text{GeV}^{-2}]$ & Fully-Dif & Single integrated & Single fine bins & Fully-Dif & Single integrated & Single fine bins & Fully-Dif & Single integrated & Single fine bins \\ \hline
    $G_{lq}^{(3)}$ & $[-2.04,\, 2.10]$ & $[-2.23,\, 2.31]$ & $[-2.07,\, 2.14]$  & $[-1.43,\, 1.47]$ & $[-1.65,\, 1.71]$ & $[-1.50,\, 1.56]$ & $[-0.73,\, 0.75]$ & $[-1.02,\, 1.08]$ & $[-0.91,\, 0.96]$\\  
    $G_{lq}^{(1)}$ & $[-8.50,\, 14.7]$ & $[-9.48,\, 20.2]$ & $[-8.85,\, 18.0]$  & $[-5.75,\, 9.01]$ & $[-6.72,\, 15.5]$ & $[-6.07,\, 12.1]$ & $[-2.66,\, 3.36]$ & $[-3.68,\, 9.22]$ & $[-3.04,\, 4.54]$\\ 
    $G_{qe}$ & $[-8.72,\, 15.2]$ & $[-11.1,\, 18.0]$ & $[-10.4,\, 16.8]$  & $[-5.98,\, 10.8]$ & $[-8.10,\, 13.9]$ & $[-7.38,\, 12.5]$ & $[-2.88,\, 5.26]$ & $[-4.68,\, 9.10]$ & $[-3.95,\, 7.19]$\\ 
    $G_{lu}$ & $[-8.21,\, 13.5]$ & $[-11.3,\, 21.1]$ & $[-10.5,\, 18.5]$  & $[-5.40,\, 8.02]$ & $[-8.03,\, 16.0]$ & $[-7.22,\, 12.7]$ & $[-2.41,\, 3.03]$ & $[-4.43,\, 10.0]$ & $[-3.65,\, 5.93]$\\ 
    $G_{ld}$ & $[-27.1,\, 18.1]$ & $[-30.5,\, 21.6]$ & $[-29.3,\, 20.5]$  & $[-20.3,\, 12.8]$ & $[-23.6,\, 16.1]$ & $[-22.3,\, 14.9]$ & $[-11.7,\, 6.70]$ & $[-15.2,\, 9.72]$ & $[-13.6,\, 8.41]$\\ 
    $G_{eu}$ & $[-6.29,\, 7.37]$ & $[-6.91,\, 8.54]$ & $[-6.27,\, 7.42]$  & $[-4.16,\, 4.64]$ & $[-4.77,\, 5.57]$ & $[-4.16,\, 4.68]$ & $[-1.88,\, 1.99]$ & $[-2.52,\, 2.75]$ & $[-2.00,\, 2.11]$\\ 
    $G_{ed}$ & $[-27.7,\, 15.5]$ & $[-29.9,\, 16.7]$ & $[-27.1,\, 15.4]$  & $[-20.5,\, 10.9]$ & $[-23.2,\, 12.2]$ & $[-19.8,\, 10.9]$ & $[-11.4,\, 5.56]$ & $[-15.4,\, 7.09]$ & $[-11.1,\, 5.84]$ \\ \hline
    \end{tabular}
    }\vspace{2mm}
    \caption{One dimensional single parameter $95\%$ confidence intervals for the seven EFT Wilson coefficients in units of $10^{-9}\, \text{GeV}^{-2}$, for integrated luminosity values of $100\,\text{fb}^{-1}$, $300\,\text{fb}^{-1}$ and $3000\,\text{fb}^{-1}$. The results are obtained considering a fully-differential cross-section, a single differential cross-section obtained from the fully-differential one integrating over angular and rapidity variables and a single differential cross-section with a fine binning.}
    \label{tab:1Dsinglepar}
\end{table*}
\begin{figure*}[t!]
    \centering
    \includegraphics[scale=0.4]{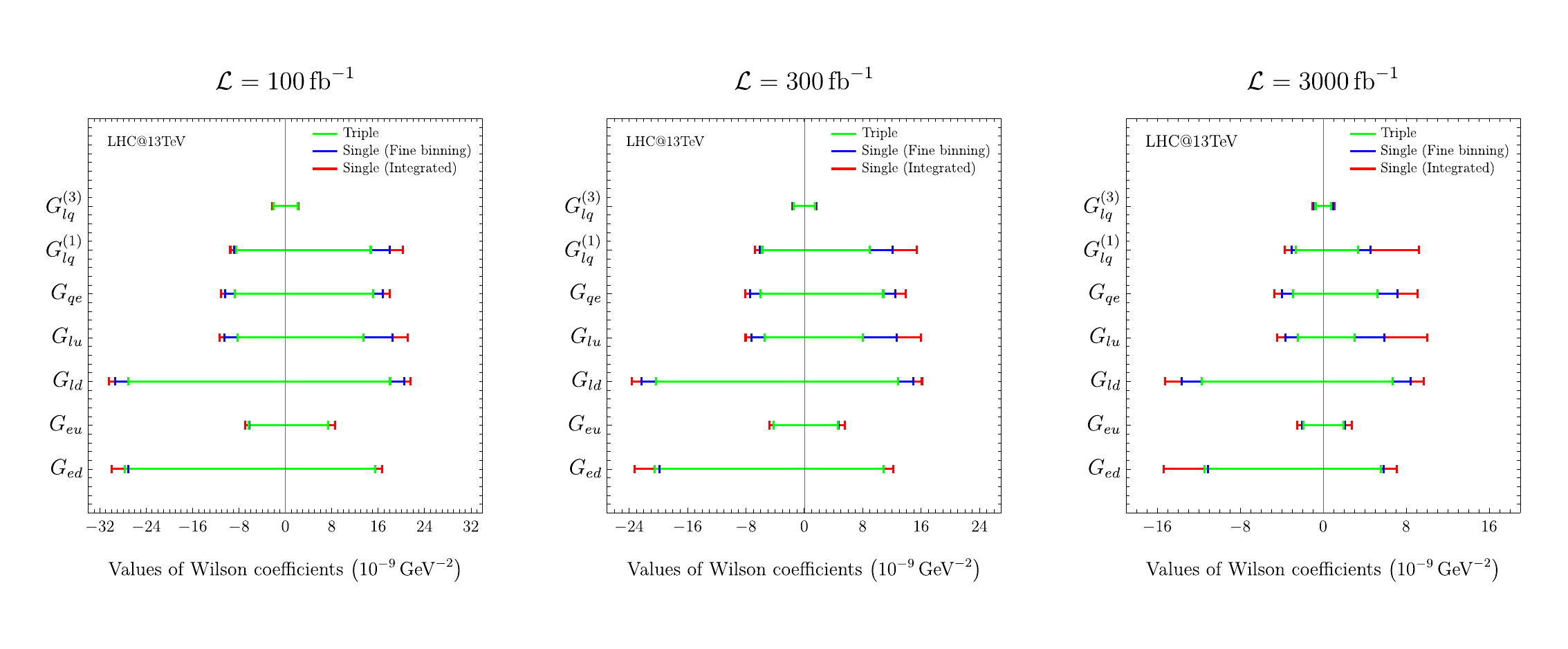}
    \caption{Comparison between the one dimensional single parameter (switch on one coefficient at a time and set the other six to 0) $95\%$ confidence intervals for the seven EFT Wilson coefficients. These results are obtained considering a multi-differential cross-section (green), a single differential cross-section obtained from the multi-differential one integrating over angular and rapidity variables (red) and a single differential cross-section with a fine binning (blue). Collider energy is set to 13 TeV and integrated luminosity of $100\,\text{fb}^{-1}$, $300\,\text{fb}^{-1}$ and $3000\,\text{fb}^{-1}$ are considered.}
    \label{fig:1Dsingleparconfront}
\end{figure*}

\section{Projections and results}
We present here the $95\%$ confidence level (CL) projected bounds, at the LHC at $13$ TeV, on each of the Wilson coefficients of the operators in Table \ref{tab:EFToperators}. We assume that the SM central value for the number of observed events in each bin is given by $n_I=L\cdot\overline{\sigma}_I$, where $L$ is the luminosity and $\overline{\sigma}_I$ is the cross-section in each bin calculated assuming the central values for $\alpha_{\text{S}}$, PDFs, and factorization and renormalization scale, and setting to zero all the experimental uncertainties and the EFT effects. We consider luminosity values of $100$, $300$, and $3000\,\text{fb}^{-1}$ as in Ref.~\cite{Torre:2020aiz}.
The binning specification for the multi-differential analysis is analogous to the one of Ref.~\cite{Panico:2021vav}. In particular, given the dilepton invariant mass $m_{\ell\ell}$, the scattering angle $c_* = \cos{\theta_*}$, and the absolute value of the ratio between the dilepton rapidity and its maximum value $\left\lvert y/y_{\text{max}}\right\rvert$, the binning for the neutral channel is given by
\begin{equation}
    \left\{
        \begin{array}{l l}
        \dst m_{\ell\ell}&: \{300, 360, 450, 600, 800, 1100,\vspace{0.5mm}\\
         & \dst  1500, 2000, 2600, 13000\}\, \text{GeV}\,,\vspace{1mm}\\
        c_*&: \dst \{-1, -0.6, -0.2, 0.2, 0.6, 1\}\,,\vspace{1mm}\\
        \dst \left\lvert\frac{y}{y_{\text{max}}}\right\rvert&:\{0, 1/3, 2/3, 1\}\,.
        \end{array}
    \right.
\end{equation}
Analogously, for the charged channel, given the lepton transverse momentum $p_T$ and the absolute value of the ratio between the lepton pseudo-rapidity and its maximum value $\left\lvert\eta/\eta_{\text{max}}\right\rvert$, the binning is given by
\begin{equation}
    \left\{
        \begin{array}{l l}
        \dst p_T&: \{150, 180, 275, 300, 400, 550,\vspace{0.5mm}\\
         & \dst  750, 1000, 1300, 7500\}\, \text{GeV}\,,\vspace{1mm}\\
        \dst \left\lvert\frac{\eta}{\eta_{\text{max}}}\right\rvert&:\{0, 1/3, 2/3, 1\}\,.
        \end{array}
    \right.
\end{equation}

\begin{table*}[t!]
    \centering
    \resizebox{\linewidth}{!}{
    \begin{tabular}{|c|c|c|c|c|c|c|c|c|c|}
    \hline \textbf{95\% CL} & \multicolumn{3}{c|}{$\mathcal{L}=100\,\text{fb}^{-1}$} & \multicolumn{3}{c|}{$\mathcal{L}=300\,\text{fb}^{-1}$} & \multicolumn{3}{c|}{$\mathcal{L}=3000\,\text{fb}^{-1}$} \\ \cline{2-10}
    $[10^{-9} \text{GeV}^{-2}]$ & Fully-Dif & Single integrated & Single fine bins & Fully-Dif & Single integrated & Single fine bins & Fully-Dif & Single integrated & Single fine bins \\ \hline
    $G_{lq}^{(3)}$ & $[-2.17,\, 2.17]$ & $[-2.35,\, 2.39]$ & $[-2.16,\, 2.18]$  & $[-1.53,\, 1.54]$ & $[-1.73,\, 1.77]$ & $[-1.57,\, 1.59]$ & $[-0.82,\, 0.82]$ & $[-1.05,\, 1.10]$ & $[-0.93,\, 0.98]$\\  
    $G_{lq}^{(1)}$ & $[-14.1,\, 17.2]$ & $[-15.7,\, 20.8]$ & $[-14.9,\, 19.5]$  & $[-10.5,\, 12.5]$ & $[-12.1,\, 16.3]$ & $[-11.2,\, 14.8]$ & $[-6.11,\, 6.78]$ & $[-7.77,\, 11.2]$ & $[-6.52,\, 9.22]$\\ 
    $G_{qe}$ & $[-12.3,\, 15.6]$ & $[-16.9,\, 18.3]$ & $[-16.1,\, 17.1]$  & $[-9.01,\, 11.5]$ & $[-13.2,\, 14.2]$ & $[-12.2,\, 12.8]$ & $[-5.12,\, 6.50]$ & $[-8.77,\, 9.36]$ & $[-7.42,\, 7.52]$\\ 
    $G_{lu}$ & $[-10.9,\, 17.4]$ & $[-18.3,\, 22.1]$ & $[-17.4,\, 20.5]$  & $[-7.56,\, 12.4]$ & $[-14.1,\, 17.0]$ & $[-13.0,\, 15.2]$ & $[-3.84,\, 6.55]$ & $[-9.11,\, 11.2]$ & $[-7.59,\, 8.81]$\\ 
    $G_{ld}$ & $[-27.9,\, 24.6]$ & $[-32.7,\, 32.0]$ & $[-30.9,\, 30.8]$  & $[-20.9,\, 18.4]$ & $[-25.7,\, 25.5]$ & $[-23.6,\, 24.1]$ & $[-12.2,\, 10.7]$ & $[-17.4,\, 17.8]$ & $[-14.6,\, 15.8]$\\ 
    $G_{eu}$ & $[-10.4,\, 17.3]$ & $[-11.8,\, 19.2]$ & $[-11.3,\, 17.6]$  & $[-7.47,\, 12.9]$ & $[-8.85,\, 14.9]$ & $[-8.30,\, 13.1]$ & $[-4.05,\, 7.75]$ & $[-5.45,\, 9.88]$ & $[-4.74,\, 7.65]$\\ 
    $G_{ed}$ & $[-28.7,\, 25.7]$ & $[-30.2,\, 28.7]$ & $[-28.1,\, 27.7]$  & $[-21.7,\, 19.5]$ & $[-23.5,\, 22.9]$ & $[-21.1,\, 21.7]$ & $[-13.1,\, 11.5]$ & $[-15.6,\, 16.0]$ & $[-12.5,\, 14.3]$ \\ \hline
    \end{tabular}
    }\vspace{2mm}
    \caption{One dimensional profiled (constraining one by one each coefficient while treating the other six as nuisance parameters) $95\%$ confidence intervals for the seven EFT Wilson coefficients in units of $10^{-9}\, \text{GeV}^{-2}$, for integrated luminosity values of $100\,\text{fb}^{-1}$, $300\,\text{fb}^{-1}$ and $3000\,\text{fb}^{-1}$. The results are obtained considering a fully-differential cross-section, a single differential cross-section obtained from the fully-differential one integrating over angular and rapidity variables and a single differential cross-section with a fine binning.}
    \label{tab:1Dprofiled}
\end{table*}
\begin{figure*}[t!]
    \centering
    \includegraphics[scale=0.4]{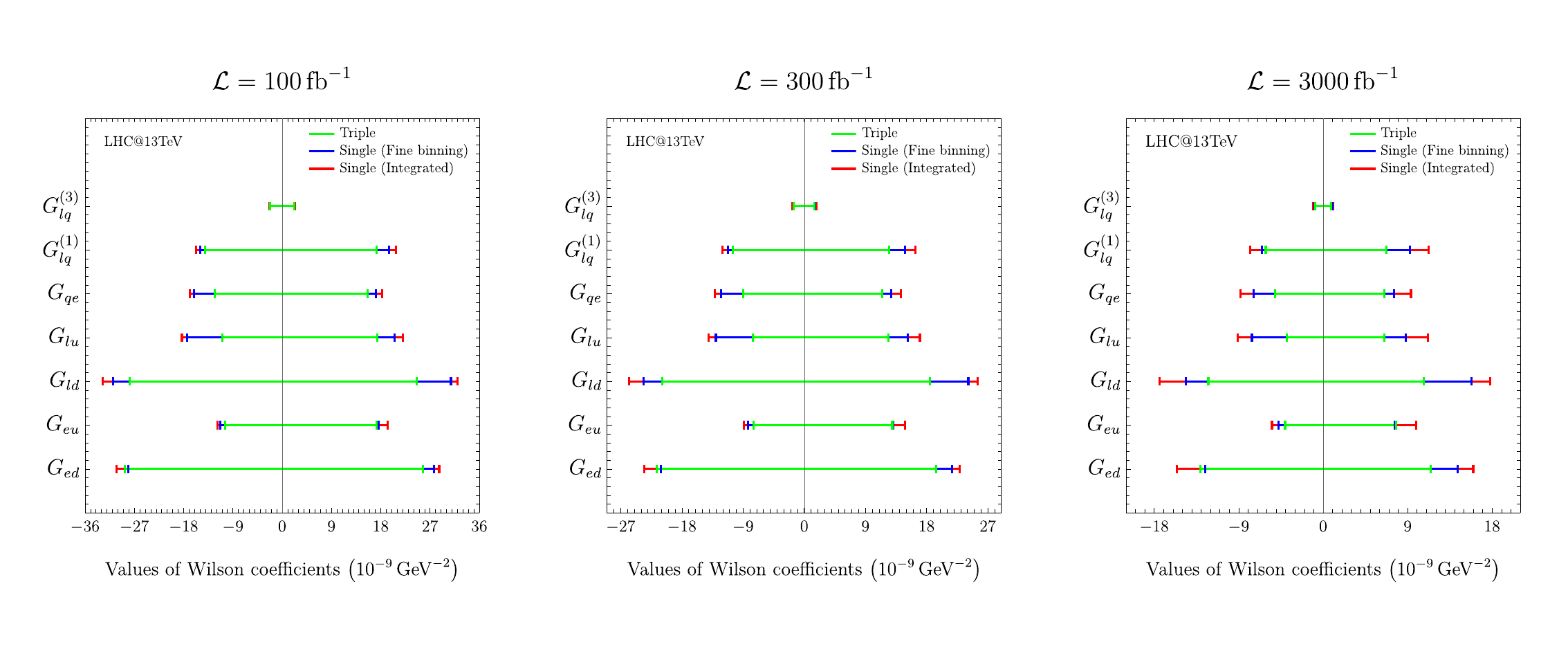}
    \caption{Comparison between the one dimensional profiled $95\%$ confidence intervals for the seven EFT Wilson coefficients. These results are obtained considering a multi-differential cross-section (green), a single differential cross-section obtained from the multi-differential one integrating over angular and rapidity variables (red) and a single differential cross-section with a fine binning (blue). Collider energy is set to 13 TeV and integrated luminosity of $100\,\text{fb}^{-1}$, $300\,\text{fb}^{-1}$ and $3000\,\text{fb}^{-1}$ are considered.}
    \label{fig:1Dsingleparconfrontprofiled}
\end{figure*}

For the single differential analysis we considered two different scenarios. In the first one the single differential cross-sections are obtained integrating the fully differential ones over angular and rapidity variables (neutral) and over pseudo-rapidity variable (charged) keeping the aforementioned binning for the dimensionful kinematic variables. In the second one the large number of expected events is exploited to consider a finer binning in the $m_{\ell\ell}$/$p_T$ (neutral/charged) variables. Such binning is obtained requiring that the statistical uncertainty of our Monte Carlo simulation remains negligible (below $0.2\%$) in all bins.
With this constraint we get $113$ bins in $m_{\ell\ell}$ for the neutral channel and $26$ bins in $p_T$ for the charged one (the exact binning is reported in Appendix~\ref{fine_binning}).\\

The constraints we obtained in the multi-differential analysis and in the single differential one after the integration over the rapidity and the angular variables are compatible with the results reported in Ref.~\cite{Panico:2021vav}. Our bounds are slightly more conservative because we also included the uncertainties deriving from the value of the strong coupling constant $\alpha_{\text{S}}$, and the truncation of the perturbative series. Furthermore, we considered a 2\% experimental uncertainty in the neutral channel and a 5\% in the charged one, while in Ref.~\cite{Panico:2021vav} a 2\% uncertainty was considered for both channels. The single differential analysis with a finer binning was not considered in previous works.\\

Table \ref{tab:1Dsinglepar} reports the $95\%$ CL projected bounds on the Wilson coefficients in the three scenarios discussed above: considering a fully-differential cross-section, a single differential cross-section obtained integrating the fully-differential one over angular and rapidity variables, and a single differential cross-section with a finer binning. The three aforementioned integrated luminosity values are considered. The numerical values in the table are obtained considering only one Wilson coefficient at a time, with all others set to zero, building the test-statistic $t_{\boldsymbol{\mu}}$ as function of a single parameter, and assuming it is asymptotically distributed as a $\chi^2$ with one degree of freedom. The results are also shown in Figure~\ref{fig:1Dsingleparconfront}, which highlights the differences between the three binning strategies.\\

From the figures and the table we can draw some important conclusions. 
\begin{itemize}
    \item As expected, the fully-differential analysis always provides the tightest constraints on the Wilson coefficients.
    \item The single-differential analysis with fine bins often reaches the same level of sensitivity as the fully-differential one. This is particularly true for the Wilson coefficients $G^{(3)}_{lq}$, $G_{eu}$, and $G_{ed}$, which are therefore much less sensitive to angular distributions. The sensitivity is still weaker than the fully-differential analysis for the other Wilson coefficients, but the difference is, especially in the negative end of the bound, generally less than $10\%$, which can be considered within the precision of our determination, based on asymptotic formulae for the likelihood-ratio test-statistic distribution.
    \item The single differential analysis obtained by integrating the fully-differential cross-section over the angular and rapidity variables is generally less sensitive than the fully-differential one. This is particularly true for the operators $G_{lq}^{(1)}$, $G_{qe}$, $G_{lu}$, and $G_{ld}$, whose upper bounds reach differences from the fully-differential analysis of more than a factor of two. This implies that the bound on these parameters is particularly sensitive to the angular and rapidity distributions.
\end{itemize}

The same results, but obtained marginalizing the likelihood over the other Wilson coefficients, instead of setting them to zero, are reported in Table \ref{tab:1Dprofiled} and shown in Figure~\ref{fig:1Dsingleparconfrontprofiled}. The conclusions are unchanged. 

\section{Conclusions}
In this work we generalized previous work on the determination of the bounds on the Wilson coefficients of the EFT operators entering in the DY processes at the LHC. We have put together the ``all-operators" approach of Ref.~\cite{Panico:2021vav} with the ``all-uncertainties" approach of Ref.~\cite{Torre:2020aiz}. We found consistent results with the previous works in the multi-differential and integrated single differential analysis~\cite{Panico:2021vav} and we added a single differential ``fine binning" analysis. In turn, we have discussed the sensitivity of different Wilson coefficients to angular and rapidity distributions, and we have shown that while the fully-differential analysis is always the most sensitive, the single differential analysis with fine bins can reach a similar level of sensitivity for a subset of parameters. This highlights the importance of presenting results in different ways, distributing the largest possible amount of information contained in the analysis.\\

\subsection*{Acknowledgments}
We thank A. Wulzer and L. Ricci for useful discussions. We also thank L. Ricci for help in comparing our results with those of Ref.~\cite{Panico:2021vav}, and E.~Rizvi and G.~Poddar for clarifications about the experimental analyses in Ref.s~\cite{ATLAS:2017rue,ATLAS:2016gic}.

\appendix

\section{Cross-section parametrization}\label{app:cross_section}

The cross-section as a function of the Wilson coefficients is a non-negative quadratic polynomial in $G^{(3)}_{lq}$, $G^{(1)}_{lq}$, $G_{qe}$, $G_{lu}$, $G_{ld}$, $G_{eu}$, $G_{ed}$. The cross-section in each bin $I$ can therefore be parametrized using the Cholesky decomposition. In the neutral channel, this can be written as:
\begin{equation}
    \sigma^{\text{th,n}}_I=\overline{\sigma}^{\text{SM,n}}_Ic^2_{0,I}\left\|\sum_{j=1}^{8}\mathcal{C}_{Ij}G_j\right\|^2\,,
\end{equation}
with the Cholesky matrix
\begin{equation}
    \mathcal{C}_{Ij}=\left(\begin{matrix}
        1 &  c_{1,I} & c_{2,I} & c_{3,I} & c_{4,I} & c_{5,I} & c_{6,I} & c_{7,I}\\
        0 &  c_{8,I} & c_{15,I} & c_{16,I} & c_{17,I} & c_{18,I} & c_{19,I} & c_{20,I}\\
        0 & 0 & c_{9,I} & c_{21,I} & c_{22,I} & c_{23,I} & c_{24,I} & c_{25,I}\\
        0 & 0 & 0 & c_{10,I} & c_{26,I} & c_{27,I} & c_{28,I} & c_{29,I}\\
        0 & 0 & 0 & 0 & c_{11,I} & c_{30,I} & c_{31,I} & c_{32,I}\\
        0 & 0 & 0 & 0 & 0 & c_{12,I} & c_{33,I} & c_{34,I}\\
        0 & 0 & 0 & 0 & 0 & 0 & c_{13,I} & c_{35,I}\\
        0 & 0 & 0 & 0 & 0 & 0 & 0 & c_{14,I}\\
    \end{matrix}\right)
\end{equation}
and the vector of Wilson coefficients $G_j$ with $j=1,...,8$
\begin{equation}\label{eq:G_j}
    G_j=\left(1, G^{(1)}_{lq}, G^{(3)}_{lq}, G_{qe}, G_{lu}, G_{ld}, G_{eu}, G_{ed}\right)\,,
\end{equation}
where the $1$ accounts for the SM contribution, and the Cholesky coefficients $c_{k,I}$ are functions of the nuisance parameters $\theta^{\alpha_{\text{S}}}$, $\theta^{\text{PDF}}_i$, and $\theta^{\text{TU}}_{I}$, as described in the main text. The same parametrization holds for the charged channel, where it is more simply written as
\begin{equation}
\sigma^{\text{th,c}}_{I}=\overline{\sigma}^{\text{SM,c}}_{I}c^2_{0,I}\left\|\left(\begin{matrix}
1 &  c_{1,I}\\
0 &  c_{2,I} 
\end{matrix}\right)\left(\begin{matrix}
1 \\
G^{(3)}_{lq}
\end{matrix}\right)\right\|^2\,.
\end{equation}

\section{Likelihood}\label{likelihood}
Each factor of the combined likelihood is:
\begin{equation*}
    \begin{array}{lll}
    \dst \mathscr{L}_{\text{n}}&=&\dst \prod_{I_{\text{n}}=1}^{N_\text{n}}\text{Poisson}\left[n_{I_{\text{n}}}\mid\mu_{I_{\text{n}}}(\mathbf{G}, \boldsymbol{\theta}_{\boldsymbol{\text{n}}})\right]\,,\vspace{2mm}\\
    \dst \mathscr{L}_{\text{c},\pm}&=&\dst \prod_{I_{\text{c}}=1}^{N_\text{c}}\text{Poisson}\left[n^{\pm}_{I_{\text{c}}}\mid\mu^{\pm}_{I_{\text{c}}}(\mathbf{G}, \boldsymbol{\theta}_{\boldsymbol{\text{c}}})\right]\,,\vspace{2mm}\\
    \end{array}
\end{equation*}
\begin{equation*}
    \begin{array}{lll}
    \small
    \dst \mathscr{L}_{\text{aux}}&=&\dst \prod_{I_{\text{n}}=1}^{N_\text{n}}\prod_{I_{\text{c}}=1}^{N_\text{c}}\prod_{i=1}^{30}f_{\alpha_{\text{S}}}(\theta^{\alpha_{\text{S}}})f_{\text{PDF}}(\theta_i^{\text{PDF}})f(\theta^{\text{TU}}_{I_{\text{n}}})\vspace{2mm}\\
    &&\dst \times f(\theta^{\text{TU}}_{I_{\text{c}}})f_{\text{exp}}(\theta^{\text{exp}}_{I_{\text{n}}})f_{\text{exp}}(\theta^{\text{exp}}_{I_{\text{c}}})f_{\text{L}}(\theta^{\text{L}})\,,
    \end{array}
\end{equation*}
where $\mathbf{G}$ is the vector defined in eq.~\eqref{eq:G_j}, and 
\begin{equation*}
    \begin{array}{lll}
    \dst \boldsymbol{\theta}_{\boldsymbol{\text{n}}} = \left(\theta^{\alpha_{\text{S}}}, \theta^{\text{PDF}}_i, \theta^{\text{TU}}_{I_{\text{n}}}, \theta^{\text{exp}}_{I_{\text{n}}}, \theta^{\text{L}}\right)\,,\vspace{2mm}\\
    \boldsymbol{\theta}_{\boldsymbol{\text{c}}} = \left(\theta^{\alpha_{\text{S}}}, \theta^{\text{PDF}}_i, \theta^{\text{TU}}_{I_{\text{c}}}, \theta^{\text{exp}}_{I_{\text{c}}}, \theta^{\text{L}}\right)\,.
    \end{array}
\end{equation*}

\section{Fine Binning}\label{fine_binning}
The ``fine binning'' have been obtained imposing to have a negligible MC statistical error in each bin (below 1\%). We achieved this imposing to have $\sim2\cdot10^{5}$ MC events for each bin in our simulation, that corresponds to a statistical error of $\sim0.22\%$. The resulting binning, for the neutral and charged channels, is:
\begin{itemize}
    \item {\it neutral} \\
    $m_{\ell\ell}$\,: \{300, 305, 309, 315, 320, 326, 332, 337, 342, 348, 355, 362, 368, 375, 381, 389, 397, 405, 413, 420, 427, 435, 444, 453, 462, 470, 478, 487, 497, 507, 518, 527, 536, 546, 556, 567, 580, 592, 602, 614, 626, 639, 653, 669, 684, 696, 709, 723, 739, 756, 774, 792, 805, 819, 834, 850, 868, 888, 908, 924, 941, 959, 979, 1001, 1026, 1053, 1077, 1095, 1115, 1137, 1161, 1188, 1218, 1251, 1275, 1297, 1322, 1349, 1379, 1413, 1451, 1494, 1521, 1548, 1577, 1610, 1647, 1689, 1737, 1792, 1823, 1854, 1888, 1926, 1969, 2018, 2076, 2142, 2184, 2220, 2259, 2305, 2357, 2417, 2489, 2575, 2680, 2910, 3105, 3365, 3752, 4126, 4802, 13000\} GeV. \vspace{1mm}
    \item {\it charged} \\
    $p_T$\,: \{150, 163, 177, 191, 207, 225, 244, 264, 288, 313, 342, 373, 407, 445, 488, 537, 591, 652, 723, 802, 896, 1003, 1130, 1292, 1493, 1770, 6500\} GeV.
\end{itemize}

\bibliographystyle{mine}
\bibliography{references}

\end{document}